\documentclass[12pt]{article}
\usepackage{amssymb}
\usepackage[mathscr]{eucal}
\usepackage{citesort}
\usepackage{url}
\usepackage{a4}
\usepackage{cite}
\usepackage{graphicx}
\usepackage{latexsym}

\newcounter{mnotecount}[section]

\begin{document}
\title{On the motion of a compact elastic body}

\author{Robert Beig\\Institut f\"ur Theoretische Physik der Universit\"at
Wien\\ Boltzmanngasse 5, A-1090 Vienna, Austria\\[1cm]
Michael Wernig-Pichler\\Institut f\"ur Theoretische Physik der Universit\"at Wien\\ Boltzmanngasse
5, A-1090 Vienna, Austria}

\maketitle

\begin{abstract}
We study the problem of motion of a relativistic, ideal elastic solid with free surface boundary by
casting the equations in material form ("Lagrangian coordinates"). By applying a basic theorem due
to Koch, we prove short-time existence and uniqueness for solutions close to a trivial solution.
This trivial, or natural, solution corresponds to a stress-free body in rigid motion.

\end{abstract}

\section{Introduction}
In the problem of motion for classical continua with free surface boundary, despite its obvious
physical relevance, there is a surprising scarcity of rigorous results. A recent review of known
results with and without gravity can be found in \cite{RE}. In the case of a nonrelativistic,
non-gravitating perfect fluid local wellposedness has only been proved very recently by Lindblad
\cite{LI}. In the present paper we consider the analogous problem for  relativistic elastic solids
without self-gravity. The nonrelativistic case of our results (which is in principle known - see
\cite{SI} - although even there a detailed treatment seems to be lacking) can be proved by similar
methods. The nonrelativistic case for incompressible materials has, by different methods, been
treated in \cite{EB}.

The plan of this paper is as follows. In Sect.2 we describe
elasticity theory on an arbitrary curved background as a Lagrangian
field theory. Sect.3 is devoted to the notion of "natural state",
that-is-to-say a solution to the elastic equations corresponding to
a configuration with zero stress. The existence of such a solution
requires the elastic body to move along a geodesic, timelike Killing
vector. We require the Killing vector to be also
hypersurface-orthogonal (thus forcing spacetime to be
"ultrastatic")\footnote{In Special Relativity a geodesic, timelike
Killing motion has to be a time translation, so the above assumption
is superfluous in that case.}. In the case where the metric on the
space orthogonal to the Killing vector is flat, we are in Special
Relativity. In Sect.4 we perform a 3+1 decomposition of the elastic
equations corresponding to the natural space-time splitting afforded
by the Killing vector. We then write the elastic equations in
"material" form (often called "Lagrangian representation"). This
means that the maps making up the elastic configuration space -
which go from spacetime to the 3 dimensional "material space" - are
replaced by time dependent maps ("deformations") from material space
into physical space. In the material representation the boundary of
the body is fixed, namely the boundary of material space. In Sect.5
we state a corollary to the theorem of Koch \cite{K}, which is the
version of existence theorem we are using. In Sect.6 we state our
basic constitutive assumption. This assumption, which is satisfied
by elastic materials occurring in practice, implies the validity of
the conditions in the Koch theorem. From this one concludes the main
result, which is stated in Sect.6. In the appendix we show that the
corner conditions on the boundary, which initial data have to
satisfy in order for the time evolved solution to be a classical
solution, can be satisfied for a large class of initial data.

We should emphasize that our requirement of the existence of a
global relaxed state, and the associated restriction on the
background spacetime would be unnecessary if we were dealing with
well-posedness in the absence of a boundary, or inside a region of
the body causally disconnected from the boundary. However in the
presence of the boundary we have to ensure corner conditions using
an implicit function argument, and for this we use the existence of
a state - the global relaxed one - which satifies these corner
conditions to all orders. In this sense our strong restriction on
the existence of a global relaxed configuration is perhaps not
completely essential, but very convenient.

We can state the central result of this paper as follows: For
initial data close to initial data for the natural deformation in
the appropriate function space, there exists, for sufficiently short
times,  a unique solution to the elastic equations. This solution
depends continuously on initial data, in particular tends to the
natural deformation, when the initial data tend to that of the
natural deformation.

\section{The theory}

We treat elasticity as a Lagrangian field theory in the manner of \cite{BS} or \cite{C}, see also
\cite{KM}. In the language of standard elasticity this means that the material in question is
"hyperelastic". The dynamical objects of the theory are furnished by sufficiently regular maps $f$
sending a closed region $\bar{\mathcal{S}}=\mathcal{S}\cup\partial\mathcal{S}$ of spacetime
$\mathcal{M}$ onto $\bar{\mathcal{B}}=\mathcal{B}\cup\partial\mathcal{B}$, with $\mathcal{B}$ a
bounded domain in $\mathbb{R}^3$ with smooth boundary $\partial \mathcal{B}$ called the "body" or
material space. The domain $\mathcal{B}$ is to be thought of as an abstract collection of points
("labels") making up the elastic continuum, and $f$ is to be thought of as the back-to-labels map
sending spacetime events to the particles by which they are occupied. We endow $\mathcal{B}$ with a
volume form $\Omega$ which is smooth on $\bar{\mathcal{B}}$. Coordinates on $\mathcal{M}$ are
written as $x^\mu$, with $\mu, \nu$ etc ranging from 0 to 3 and coordinates on $\mathcal{B}$ as
$X^A$ with $A,B$ etc going from 1 to 3. The metric $g_{\mu\nu}$ is also taken to be smooth. There
are additional requirements on the maps $f$. Namely the equation
\begin{equation} \label{unit}
f^A{}_{,\mu} u^\mu = 0, \qquad g_{\mu\nu} u^\mu u^\nu = -1
\end{equation}
should have a solution $u^\mu$ which is unique up to sign. Thus particles move along trajectories
in $\mathcal{M}$ with unit tangent $u^\mu$. Furthermore the function $n$ on $\bar{\mathcal{S}}$
defined by the equation
\begin{equation}\label{n}
\Omega_{ABC}(f(x))
f^A{}_{,\mu}(x)f^B{}_{,\nu}(x)f^C{}_{,\lambda}(x) = n(x)
\epsilon_{\mu\nu\lambda\rho}(x) u^\rho(x)
\end{equation}
should be everywhere positive on $\bar{\mathcal{S}}$.
It is well known that (\ref{n}) implies that the
continuity equation
\begin{equation}\label{cont}
\nabla_\mu (n u^\mu) = 0
\end{equation}
is identically satisfied. The equations of motion are the
Euler-Lagrange equations of the action

\begin{equation}\label{action}
S[f,\partial f] = \int_{\mathcal{S}} \rho (f,\partial f,x)\, (-g)^{\frac{1}{2}}\; d^4x .
\end{equation}
The function $\rho$ is required to be smooth in all its arguments.
It plays the double role of being the Lagrangian as well as energy
(i.e. rest mass plus internal elastic energy) density of the
material. Diffeomorphism covariance demands that $\rho$ be of the
form

\begin{equation}\label{diffeo}
\rho (f, \partial f,x) = \hat{\rho} (f,H^{AB})|_{H^{AB}=h^{AB}(\partial f,x)},
\end{equation}
where $h^{AB}(\partial f,x) = f^A{}_{,\mu}(x)f^B{}_{,\nu}(x)g^{\mu\nu}(x)$. Thus $\rho$ depends
only on points $X$ on $\mathcal{B}$ and positive definite contravariant tensors $H^{AB}$ (see
\cite{BS}). (In nonrelativistic elasticity materials governed by such a Lagrangian are said to
satisfy the condition of "material frame indifference".) By abuse of notation we henceforth omit
the hat from $\hat{\rho}$. It is useful to know that the Euler-Lagrange equations are equivalent to
divergence-lessness of the symmetric energy momentum tensor. More precisely, there holds the
identity (see \cite{BS})

\begin{equation}\label{ident}
-\nabla_\nu T_\mu{}^\nu = \mathscr{E}_A f^A{}_{,\mu},
\end{equation}
where
\begin{equation}\label{sym}
T_{\mu \nu} = 2 \frac{\partial \rho}{\partial g^{\mu \nu}} - \rho
g_{\mu \nu}
\end{equation}
and
\begin{equation}\label{euler}
- \mathscr{E}_A = (-g)^{-\frac{1}{2}}\partial_\mu \left(
(-g)^{\frac{1}{2}} \frac{\partial \rho}{\partial(f^A{}_{,\mu})}
\right) - \frac{\partial\rho}{\partial f^A}.
\end{equation}
The relation (\ref{ident}) shows that of the four conservation laws
$\nabla_\nu T_\mu{}^\nu = 0$ only three are independent. This, of
course, is due to the fact that within the present formalism the
continuity equation is an identity.

We now turn to boundary conditions. The boundary conditions usually
considered in nonrelativistic elasticity are either the so-called
"boundary conditions of place" - where, in our language, the set
$f^{-1}(\partial \mathcal{B})$ is prescribed, or "traction boundary
conditions" - where the normal traction, i.e. the components of the
stress tensor normal to this surface are prescribed. When the normal
traction is zero, one speaks of "natural" boundary conditions: these
are the boundary conditions appropriate for a freely floating
elastic body considered in the present work. They are employed e.g.
in geophysics for the elastic motion corresponding to seismic waves,
the free boundary in question corresponding to the surface of the
earth (see \cite{AR} \footnote{Needless to say, these studies are
confined to the linear approximation in which the problem of the
free boundary disappears.}).While one could in the present framework
in principle consider all the above boundary conditions, it is
interesting to observe that these conditions - except for the
natural condition - become inconsistent once one couples to the
Einstein equations. Namely the standard junction conditions,
together with the Einstein equations, imply that $T_\mu{}^\nu
n_\nu|_{f^{-1}(\partial \mathcal{B})}$ be zero, where $n_\mu$ is the
conormal of $f^{-1}(\partial \mathcal{B})$. However these conditions
are precisely the natural boundary conditions. To see this, we have
to first compute the right hand side of Eq.(\ref{sym}). The result
is
\begin{equation} \label{t}
T_{\mu \nu} = \rho u_\mu u_\nu + t_{\mu \nu},
\end{equation}
where $t_{\mu \nu}$ is the (negative) Cauchy stress tensor given by
\begin{equation} \label{cauchy}
 t_{\mu \nu} = n
\tau_{AB}f^A{}_{,\mu} f^B{}_{,\nu}
\end{equation}
and we have written $\rho$ as
\begin{equation}\label{rho}
\rho = n \epsilon
\end{equation}\label{tau}
and $\tau_{AB} = 2\,\frac{\partial \epsilon}{\partial H^{AB}}$. Note this makes sense, since $n$ -
whence $\epsilon$ - is a function purely of $f$ and $H^{AB}$, as is apparent from the identity
\begin{equation}\label{sense}
6 n^2 = H^{AA'}H^{BB'}H^{CC'}\Omega_{ABC}\Omega_{A'B'C'}.
\end{equation}
The function $\epsilon$ is the relativistic version of the "stored-energy-function" of standard
elasticity. The quantity $\tau_{AB}$ corresponds to the negative of the "second Piola-Kirchhoff
stress" of nonrelativistic elasticity. Now to the boundary conditions, namely
\begin{equation} \label{bound}
T_\mu{}^\nu n_\nu|_{f^{-1}(\partial \mathcal{B})} = 0.
\end{equation}
It follows from the tangency of $u^\mu$ to the inverse images of
points of $\mathcal{B}$ under $f$, that $(u,n) = 0$. Consequently
the contraction of Eq.(\ref{bound}) with $u^\mu$, using (\ref{t}),
is identically satisfied. The remaining components yield
\begin{equation}\label{remaining}
\tau_{AB}f^B{}_{,\mu} g^{\mu\nu}n_\nu |_{f^{-1}(\partial
\mathcal{B})} = 0
\end{equation}
Equation (\ref{remaining}) will turn out to be a Neumann-type boundary condition, but it has a free
(i.e. determined-by-$f$) boundary.

\section{Natural configuration}

We assume there exists a contravariant
metric $H_0^{AB}(X)$, smooth and positive definite on
$\bar{\mathcal{B}}$, such that
\begin{equation}\label{natural}
\epsilon|_{(X,H^{AB}=H_0^{AB}(X))} = \epsilon_0 (X) > 0,
\hspace{1cm} \frac{\partial \epsilon}{\partial
H^{CD}}\Bigg{|}_{(X,H^{AB}=H_0^{AB}(X))} = 0
\end{equation}
on $\bar{\mathcal{B}}$. The quantity $H_0^{AB}$ measures local
distances in the relaxed body. The function $\epsilon_0(X)$ is the
density of rest mass in the stressfree configuration.

If there exists a configuration $f_0$ such that
\begin{equation}\label{natural1}
H_0^{AB}(f_0(x)) = h^{AB}(\partial f_{0}(x), x)= f_0^A{}_{,\mu}(x)f_0^B{}_{,\nu}(x)g^{\mu\nu}(x)=:h_{0}^{AB}(x),
\end{equation}
this map $f_0$ is called a natural or relaxed configuration. Note that spacetime has to be special
in order for a natural configuration to exist. Namely it follows from (\ref{natural1}) that
$h^{AB}_0$ has to be constant along $u_0^\mu$, the four velocity associated with $f_0$, from which
one infers that $u_0^\mu$ is a Born rigid motion (see\cite{TR}), i.e.
\begin{equation}\label{born}
\mathcal{L}_{u_0} (g_{\mu\nu} + u_{0\,\mu} u_{0\,\nu}) = 0
\end{equation}
One deduces from (\ref{ident},\ref{sym},\ref{t}) and (\ref{natural}) that $f_0$ solves the
equations of motion $\mathscr{E}_A = 0$ if and only if this rigid motion is geodesic. This in turn
implies that $u_0^\mu$ is Killing. In this work we will assume that $u_0^\mu$ is in addition
irrotational. Using coordinates in which $u_0^\mu
\partial_\mu =
\partial_t$ and $u_{0\,\mu} dx^\mu = -dt$, the spacetime metric has thus to be of the form
($i,j= 1,2,3$)
\begin{equation}\label{metric}
g_{\mu\nu}dx^\mu dx^\nu = - dt^2 + g_{ij}(x^k) dx^i dx^j.
\end{equation}
(In particular: when the positive definite three metric $g_{ij}$ is flat, we are in
 Minkowski spacetime.) With this choice of coordinates, the natural configuration
$f_0$ is of the form $f_0(t,x) = f_0(x)$ with $f_0(x)$ a smooth, invertible function
$\bar{\mathcal{S}} \cap \{t=0\} \rightarrow \bar{\mathcal{B}}$ where, using coordinates on $M$ and
$\bar{\mathcal{B}}$ in which $\epsilon_{ijk}$ and $\Omega_{ABC}$ are both positive,
$det(f_0^A{}_{,i})$ is positive on $\bar{\mathcal{S}} \cap \{t=0\}$, the latter condition
guaranteeing that $n_0$ is positive and $h_0^{AB}$ is positive definite on $\bar{\mathcal{S}} \cap
\{t=0\}$. The inverse of $f_0(x)$ will be denoted by $\Phi_0(X)$.
\section{3 + 1 and the material picture}
In this section we switch to the material description of the elastic medium by changing to
``Lagrangian'' coordinates. In a first step we replace the four velocity $u^\mu$ corresponding to
the configuration $f$ by the coordinate velocity $v^\mu \partial_\mu = \partial_t + V^i
\partial_i$, which is a multiple of $u^{\mu}$. Using (\ref{unit}) we find that $V^i$ is given in
terms of $f^A$ by (note that $f^{A}{}_{,i}$ is non-degenerate by construction)
\begin{equation}\label{V}
\dot{f}^A + f^A{}_{,i} V^i = 0,
\end{equation}
where a dot denotes partial differentiation w.r. to $t$. It follows from (\ref{metric}) that
\begin{equation}\label{main}
h^{AB} = f^A{}_{,\mu} f^B{}_{,\nu} g^{\mu\nu} = f^A{}_{,i} f^B{}_{,j}\, (g^{ij} - V^i V^j).
\end{equation}
The timelike nature of $u^\mu$ (equivalently: the positive
definiteness of $h^{AB}$) requires that
\begin{equation}\label{timelike}
|V|^2 = g_{ij}V^i V^j < 1
\end{equation}
We also see that
\begin{equation}\label{det k}
n = k\,(1 - |V|^2)^{1/2},
\end{equation}
where $k$ is defined by
\begin{equation}\label{detk1}
k \:\epsilon_{ijk} = f^A{}_{,i}f^B{}_{,j}f^C{}_{,k}\,\Omega_{ABC}
\end{equation}
with $\epsilon_{ijk}$ being the volume element of $g_{ij}$. Consequently the action (\ref{action})
takes the form
\begin{equation}\label{action1}
S = \int_{\mathcal{S}} (1 - |V|^2)^{1/2}\,\epsilon \,k\,(\det\,g_{ij})^{1/2} dt\, d^3x
\end{equation}
We now pass over to the material representation. By this we mean that configurations $f^A$ are
replaced by "deformations" $x^i = \Phi^i(t,X)$ defined by
\begin{equation}\label{deform}
f^A(t,\Phi(t,X)))= X^A
\end{equation}
with $det(\Phi^i{}_{,A})$ positive in $\bar{\mathcal{B}}$.
 It follows from (\ref{deform}) that
\begin{equation}\label{def1}
f^A{}_{,i}(t,\Phi(t,X))\, \Phi^i{}_{,B}(t,X) = \delta ^A{}_B,\hspace{0.4cm}  \Phi^i{}_{,A}(t,X)\,
f^A{}_{,j}(t,\Phi(t,X))\,= \delta ^i{}_j
\end{equation}
and
\begin{equation}\label{def2}
\dot{f}^A(t,\Phi(t,X)) + f^A{}_{,i}(t,\Phi(t,X))\dot{\Phi}^i(t,X)=0
\end{equation}
and, from (\ref{def2}) and (\ref{V}), that
\begin{equation}\label{vel}
V^i(t,\Phi(t,X)) = \dot{\Phi}^i(t,X)
\end{equation}
In order to study the field equations in the material
representation it will be extremely convenient to change
representation directly in the action (\ref{action1}). Using
(\ref{detk1}) there holds
\begin{equation}\label{action2}
S = \int_{\{ t \}\times \mathcal{B}} \epsilon \,\gamma^{-1} \, dt\,
d^3\Omega
\end{equation}
with
\begin{equation}\label{gamma}
\gamma^{-1} = (1 - |V|^2)^{\frac{1}{2}}
\end{equation}
and $d^3\Omega = \Omega_{ABC}(X)\, dX^A \wedge dX^B \wedge dX^C =
\Omega^{\frac{1}{2}}\,
 d^3X$. Now the field equations take the form
\begin{equation}\label{field}
\partial_t \!\left(\frac{\partial L}{\partial \dot{\Phi}^i}\right) + \partial_A \!\left(\frac{\partial L}{\partial
\Phi^i{}_{,A}}\right) - \frac{\partial L}{\partial \Phi^i} = 0
\hspace{0.3cm}\mathrm{in}\;\mathcal{B}
\end{equation}
where $L = \epsilon \,\gamma^{-1} \,\Omega^{\frac{1}{2}}$ for functions $\Phi$ such that
$det(\Phi^i{}_A)$ is positive in $\bar{\mathcal{B}}$. Note that $\Omega$ depends only on $X^A$,
$g_{ij}$ depends only on $\Phi^i$ and $\gamma^{-1}$ depends only on $(\Phi^i,\dot{\Phi}^j)$.
Finally $h^{AB}$, i.e.
\begin{equation}\label{hagain}
h^{AB} = F^A{}_i (\Phi^k{}_{,C}) F^B{}_j(\Phi^l{}_{,D}) [g^{ij}(\Phi) - \dot{\Phi}^i\dot{\Phi}^j],
\end{equation}
where $F^A{}_i(\Phi^j{}_{,B})$ is the inverse of $\Phi^i{}_{,A}$, depends on
$(\Phi^i,\dot{\Phi}^j,\Phi^k{}_{,A})$, so that $\epsilon$ depends on  $(X^A,\Phi^i,
\dot{\Phi}^j,\Phi^k{}_{,B})$. Using these facts, together with the identity
\begin{equation}\label{ident1}
\frac{\partial h^{AB}}{\partial \Phi^i{}_{,C}} = - 2 h^{C(A}F^{B)}{}_i,
\end{equation}
we find that
\begin{equation}\label{ij}
\frac{\partial L}{\partial \Phi^i{}_{,A}}= - 2\,
\Omega^{\frac{1}{2}} N \gamma ^{-1} \frac{\partial
\epsilon}{\partial H^{BC}}\Bigg{|}_{H^{DE}=h^{DE}}\, h^{AB}F^C{}_i
\end{equation}
and
\begin{equation}\label{tt}
\frac{\partial L}{\partial \dot{\Phi}^i} = - \Omega^{\frac{1}{2}}
\left[\epsilon\, \gamma\, g_{ij} + 2 \left(\frac{\partial
\epsilon}{\partial H^{AB}}\right)\Bigg{|}_{H^{CD}=h^{CD}} F^A{}_i
F^B{}_j\right] \dot{\Phi}^j.
\end{equation}
From (\ref{ij}) we infer that the boundary conditions (\ref{remaining}) are simply equivalent to
\begin{equation}\label{matbound}
\left(\frac{\partial L}{\partial
\Phi^i{}_{,A}}\right)n_A\Bigg{|}_{\{t\}\times \partial \mathcal{B}}
= 0,
\end{equation}
where $n_A$ is a conormal of $\partial \mathcal{B}$.

We now turn to the natural deformation, which is the material version of the natural configuration
described in Sect.3. In the chosen foliation it is time-independent, namely of the form
\begin{equation}\label{new}
\Phi^i(t,X)= \Phi_0^i(X),
\end{equation}
where $\Phi_0 (f_0 (x)) = x$. Furthermore we have that
\begin{equation}\label{initial}
F^A{}_i (\Phi_0{}^k{}_{,C})
F^B{}_j(\Phi_0{}^l{}_{,D})g^{ij}(\Phi_0(X)) = H_0^{AB}(X),
\end{equation}
in short: $F_0{}^A{}_i\, F_0{}^B{}_j\, g_0{}^{ij}=H_0^{AB}$. We know from Sect.3 that the field
equations and the boundary conditions are identically satisfied in the stress-free state. In the
present context this fact takes the form
\begin{equation}\label{new1}
 \left(\frac{\partial L}{\partial
\dot{\Phi}^i}\right)_{\!\!0}=0,\hspace{0.5cm}\left(\frac{\partial
L}{\partial
\Phi^i{}_{,A}}\right)_{\!\!0}=0,\hspace{0.8cm}\left(\frac{\partial
L}{\partial \Phi^i}\right)_{\!\!0}=0,
\end{equation}
where the subscript $0$ means evaluation in the stress-free state
$(\Phi^i, \dot{\Phi}^i)=(\Phi_0{}^i,0)$. To derive Eq.(\ref{new1})
we have used (\ref{natural},\ref{ij},\ref{tt}).

The causality properties of this system are essentially governed by the nature of the coefficients
$M^{tt}{}_{ij}$ of $\stackrel{..}{\Phi}^j$ and $M^{AB}{}_{ij}$ of $\Phi^j{}_{,AB}$. In order to be
able to apply the theorem of Koch \cite{K}, we will need a "negativity" condition on
$M^{tt}{}_{ij}$ and a certain positivity ("coerciveness") condition on $M^{AB}{}_{ij}$. We easily
find that
\begin{equation}\label{Mtt}
M_0^{tt}{}_{ij}=\left(\frac{\partial^2 L}{\partial \dot{\Phi}^i
\partial \dot{\Phi}^j}\right)
_{\!\!0}= - \Omega^{\frac{1}{2}}\,\epsilon_0\, g_{0\,ij}
\end{equation}
 For the coefficient of $\Phi^j{}_{,AB}$, using (\ref{ij}), observe that
\begin{equation}\label{Mij}
M_0^{AB}{}_{ij}=\left(\frac{\partial^2 L}{\partial \Phi^i{}_{,A}\partial
\Phi^j{}_{,B}}\right)_{\!\!0} = 4\, \Omega^{\frac{1}{2}}\, L_{0\,CEDF}\,
H_0^{AE}H_0^{BF}F_0{}^C{}_i\, F_0{}^D{}_j,
\end{equation}
where
\begin{equation}\label{M}
L_{0\,ABCD} := \left(\frac{\partial^2 \epsilon}{\partial
H^{AB}\partial H^{CD}}\right)\Bigg{|}_{H^{EF} = H_0^{EF}}
\end{equation}

\section{Koch theorem}
For convenience we state here a corollary of the theorem in \cite{K}, which is the precise
statement we need.\\

Let $\Phi^i(t,X)$ be maps from ${\{t\}\times \mathcal{B}}$ to
$\mathbb{R}^3$, where $\mathcal{B}$ is an open set in
$\mathbb{R}^3$ with smooth boundary $\partial \mathcal{B}$. We are
given a system of the form of
\begin{equation}\label{form}
\partial_\alpha \mathscr{F}^\alpha{}_i = w_i \hspace{0.3cm}\mathrm{in} \;\{t\}\times \mathcal{B},
\end{equation}
where $\alpha$ runs from 0 to 3 and with $\mathscr{F}^\alpha{}_i$ and $w_j$ all being smooth
functions of $(X^A,\Phi^i, \Phi^j{}_{,\alpha})$ on $\bar{\mathcal{B}} \times \mathbb{R}^3 \times
\mathbb{R}^{12}$, together with the boundary conditions
\begin{equation}\label{bound2}
\mathscr{F}^{\alpha}{}_{i} n_{\alpha}|_{\{t\}\times \partial \mathcal{B}}=\mathscr{F}^A{}_i
n_A|_{\{t\}\times \partial \mathcal{B}}=0.
\end{equation}

We make the following further assumptions:

(i) Symmetry: there hold the symmetries $M^{\alpha \beta}{}_{ij} = M^{\beta \alpha}{}_{ji}$ where
$M^{\alpha \beta}{}_{ij} = \frac{\partial \mathscr{F}^\alpha{}_i}{\partial \Phi^j{}_{,\beta}}$.

(ii) Static solution: There is given a time independent function $\Phi_0^i(t,X)=\Phi_0^i(X)\in
C^\infty(\bar{\mathcal{B}})$ satisfying
\begin{equation} \label{reference}
\mathscr{F}^\alpha_{0\,i} = 0,\hspace{0.5cm}w_{0\;i}=0
\end{equation}
and the following estimates

(iii) Time components:
\begin{equation}\label{1}
M_0^{tt}{}_{ij}\eta ^i \eta^j \leq -\kappa |\eta|^2
\hspace{0.4cm}\mathrm{in}\; \mathcal{B}
\end{equation}
for a positive constant $\kappa$,

(iv) Space components:
\begin{equation}\label{2}
\int_{\mathcal{B}} M_0^{AB}{}_{ij}\, \, \delta
\Phi^i{}_{,A}\,\delta \Phi^j{}_{,B}\,d^3X + ||\delta \Phi
||^2_{L^2(\mathcal{B})} \geq \sigma ||\delta
\Phi||^2_{H^1(\mathcal{B})}
\end{equation}
for $\sigma > 0$ and all $\delta \Phi \in
C^\infty(\bar{\mathcal{B}})$, where $H^s$ denotes the Sobolev
space $H^{s,2}$. We remark that condition (\ref{2}) implies the
"strong ellipticity" or "Legendre-Hadamard" condition, namely that
$M_0^{AB}{}_{ij}N_A N_B k^i k^j > 0$ in $\bar{\mathcal{B}}$ for
non-zero $N_A,k^i$. This latter condition is the relevant one for
wellposedness in the pure initial value problem (see \cite{HKM}).

{\bf Theorem:} Let $(\Phi(0),\dot{\Phi}(0))$ lie in a small neighborhood of $(\Phi_0,0)$ in
$H^{s+1}(\mathcal{B})\times H^s(\mathcal{B}), s\geq3$ and satisfy the corner conditions of order
$s$ in the sense that $\partial_t^r \mathscr{F}^A_i n_A|_{\{t=0\}\times \partial \mathcal{B}}$ is
in $H^{s-r}(\mathcal{B}) \cap H_0^1(\mathcal{B})$ for $0 \leq r \leq s$ \footnote{This condition
says that $(\partial_t^r \mathscr{F}^A{}_i)\, n_A|_{\{t=0\}\times \partial \mathcal{B}}$ is zero for $0
\leq r \leq s$ In the appendix we show that it can be met by a suitable choice of odd-order normal
derivatives of $(\Phi(0),\dot{\Phi}(0))$ on $\partial \mathcal{B}$.}. Then there exists, for
sufficiently small $t_{0}$, a unique classical solution $\Phi$ of (\ref{form},\ref{bound2}) in
$C^2([0,t_0)\times \bar{\mathcal{B}})$ with $(\Phi(0) ,\dot \Phi(0) )$ as initial data. Furthermore
$\partial^r \Phi(t) \in L^2(\mathcal{B})$ for $0 \leq r \leq s+1$. In the last expression
$\partial^r \Phi$ denotes all partial derivatives of order $r$. The evolved solution stays close to
the static solution $\Phi_0$ in the sense that $(\Phi(t),\dot{\Phi}(t))$ remains close to
$(\Phi_0,0)$ in $H^{s+1}(\mathcal{B}) \times H^s(\mathcal{B})$ and $\partial^r \Phi(t)$ remains
close to $\partial^r \Phi_0 \in L^2(\mathcal{B})$ for $0 \leq r \leq s+1$. (This last statement is
the "stability" part of the theorem.)

\section{Main result}
We now add our final constitutive assumption, sometimes called
"pointwise stability" see \cite{MH}, namely that
\begin{equation}\label{pointwise}
L_{0\,ABCD}\, N^{AB}N^{CD} \geq \sigma \,(H_{0\,CA}H_{0\,BD}+ H_{0\,CB}H_{0\,AD})\,N^{AB}N^{CD}\;\;\;\mathrm{in}\;\mathcal{B}
\end{equation}
where $\sigma$ is a positive constant which only depends on the
choice of coordinates and with $H_{0\,AB}$ being the inverse of
$H_0^{AB}$. Condition (\ref{pointwise}) is a convexity condition on
$\rho$ as a function of $H^{AB}$. We note that this in general does
not imply convexity with respect to the deformation gradient
$\Phi^i{}_{,A}$ (see \cite{CI}). An important special case is where
\begin{equation}\label{isotropic}
4 \epsilon_0 L_{0\,ABCD} = \lambda H_{0\,AB} H_{0\,CD} + \mu (H_{0\,CA} H_{0\,BD} +
H_{0\,CB}H_{0\,AD})
\end{equation}
and
\begin{equation}\label{lame} \mu (X)> 0,\hspace{0.4cm} 3 \lambda(X) + 2
\mu(X) > 0 \;\;\;\mathrm{in}\;\bar{\mathcal{B}}
\end{equation}
The quantities $\mu$ and $\lambda$, when they are independent of
$X$, are the Lam\'e constants of homogeneous, isotropic materials.
We now invoke the Korn inequality (see e.g. \cite{HO} or \cite{D})
of which we need a slight generalization due to \cite{CJ}, in the
following form: Let $\Psi^A$ be vector field on
$(\mathcal{B},H_{0\,AB})$ in some chart. Let $L(\Psi)$ be defined by
$L(\Psi)_{AB} = 2 H_{0\,C(A} \Psi^C{}_{,B)}$, i.e. the Lie
derivative of $H_{0\,AB}$ w.r. to $\Psi$ to leading derivative-order
in $\Psi^A$. Then there is a positive constant $\sigma'$ such that
\begin{equation}\label{korn}
\Vert L(\Psi)\Vert^2_{L^2(\mathcal{B})}  + \Vert \Psi \Vert^2_{L^2(\mathcal{B})} \geq \sigma' \Vert
\Psi \Vert^2_{H^1(\mathcal{B})}.
\end{equation}
We now assume coordinates $X$ in $\mathcal{B}$ to be chosen so that $\Phi_0$ is the identity map.
Combining (\ref{Mij}) with (\ref{pointwise}) and using (\ref{korn}), there follows condition (iv),
i.e. that
\begin{equation}\label{coercive}
\int_{\mathcal{B}} M_0^{AB}{}_{ij}\, \delta\Phi^i{}_{,A} \delta\Phi^j{}_{,B} ´\,d^3X + \Vert
\delta\Phi \Vert^2_{L^2(\mathcal{B})} \geq \sigma'' \Vert\delta\Phi\Vert^2_{H^1(\mathcal{B})}
\end{equation}
for positive $\sigma''$. The validity of the condition (iii) in
the Koch theorem is immediate from Eq.(\ref{Mtt}). Furthermore the
validity of (ii) has been checked in Sect.4. The validity of (i)
is obvious from the variational character of the equations. Thus
all assumptions of the Koch theorem are met.

It remains to check the determinant condition $det(\Phi^i{}_{,A})
> 0$ in $\bar{\mathcal{B}}$. But, when $(\Phi(0),\dot{\Phi}(0))$
is close to $(\Phi_0,0)$ in $H^{s+1}(\mathcal{B})\times
H^s(\mathcal{B}), s\geq3$, this immediately follows by Sobolev
embedding and the positivity of $det(\Phi_0^i{}_{,A})$ in
$\bar{\mathcal{B}}$. Now the precise form of our final statement
can be read off from the Koch theorem in Sect.5.

Stated somewhat informally, our final result is as follows:\\

{\bf Theorem}: Let there be given a volume form $\Omega(X^A)$, a
spacetime metric $g_{\mu\nu}(x^\lambda)$ of the form (\ref{metric})
and an internal energy $\epsilon(X^A,H^{BC})$ satisfying
(\ref{natural}), all smooth functions of their respective arguments.
Suppose there exists a smooth natural (i.e. stress-less)
configuration such that the corresponding vector field $u_{0}^{\mu}$
is the static Killing field $\partial_t$. Also suppose that the
elasticity tensor for this natural configuration satisfies the
pointwise stability condition (\ref{pointwise}). Let the initial
data $(\Phi(0),\dot{\Phi}(0))$ for the deformation be close to those
for the natural deformation and satisfy the corner condition of the
appropriate order. Then there exists, for these initial data, a
solution $\Phi(t)$ of the dynamical equations (\ref{field}) with
boundary conditions (\ref{matbound}) for small
times. This solution remains close to the natural deformation.\\

 We end with a remark on the possible generalization of
the results presented here to the case of an elastic body (or
bodies, if there are several) which are self-gravitating. The
presence in G.R. of constraints and gauge freedom will then make
things more complicated. Furthermore the fact that the elastic
equations are written in material form, but the Einstein equations
are equations on spacetime, means that one is now not dealing with a
system of partial differential equations in the strict sense.\\

{\bf{Acknowledgement}}: We are grateful to Bernd Schmidt for very useful remarks on the manuscript.
One of us (R.B.) thanks Tom Sideris for pointing out the relevance of the work of Koch and Sergio
Dain for helpful discussions on the Korn inequality. This work has been supported by Fonds zur
F\"orderung der Wissenschaftlichen Forschung in \"Osterreich, project no. P16745-NO2

\section{Appendix: corner conditions}

We study corner conditions for the system (\ref{form}) with the
boundary conditions (\ref{bound2}). Suppose that
$(\Phi(0)|_{\partial\mathcal{B}},
\dot{\Phi}(0)|_{\partial\mathcal{B}})$ is sufficiently close to
$(\Phi_0|_{\partial\mathcal{B}},0)$. In order for obtaining a
solution of the equations of motion, one has to be able to satisfy
conditions on the behaviour of certain normal derivatives of
$(\Phi(0), \dot{\Phi}(0))$ on $\partial\mathcal{B}$, which guarantee
that, not only $\mathscr{F}^A{}_i\, n_{A}|_{\partial\mathcal{B}}=0$
at $t = 0$, but also a sufficient number of time derivatives at $t =
0$ of that condition is satisfied. (It is understood that
higher-than-second-order time derivatives are eliminated in terms of
spatial and lower-order time derivatives, using the equations of
motion: this by virtue of the negativity of $M_0^{tt}{}_{ij}$ is
always possible for $(\Phi(0)|_{\partial\mathcal{B}},
\dot{\Phi}(0)|_{\partial\mathcal{B}})$ close to
$(\Phi_0|_{\partial\mathcal{B}}, 0)$.)

First recall that $\Phi_0(t,X)= \Phi_0(X)$ identically satisfies
the equations of motion and the boundary condition. Now we check
that the corner condition of order 0, i.e. the
undifferentiated-in-time Eq.(\ref{matbound}) can be satisfied by a
choice of $\partial_n \Phi(0)|_{\partial \mathcal{B}}$ (where
$\partial_n$ denotes any derivative transversal to
$\partial\mathcal{B}$). To see this notice first that, by the above
observation, $(\Phi_0|_{\partial\mathcal{B}},0)$ solves the
order-0 corner condition. Furthermore
\begin{equation}\label{finite}
\frac{\partial^2 L}{\partial \partial_n \Phi^j \partial
\Phi^i{}_{,A}} n_A \Bigg{|}_{\partial \mathcal{B}} =
\frac{\partial^2 L}{\partial \Phi^j{}_{,B} \partial \Phi^i{}_{,A}}
n_A n_B \Bigg{|}_{\partial \mathcal{B}}.
\end{equation}
Then the result follows from the
(finite-dimensional) implicit function theorem using that,
by virtue of (\ref{pointwise}), the quadratic form
\begin{equation}\label{nonsing}
M_0^{AB}{}_{ij}\;n_A n_B|_ {\partial\mathcal{B}}
\end{equation}
is nonsingular, which in turn follows from the strong ellipticity
condition (see Sect.5). One now performs the process of taking
($s\geq 1$ say) time derivatives of the boundary condition and
eliminating $\partial_t^m \Phi(0)|_{\partial\mathcal{B}}$ for $s+1
\ge m> 1$. Using that $\mathscr{F}^0{}_i$ is zero for
$(\Phi(0)|_{\partial\mathcal{B}},
\dot{\Phi}(0)|_{\partial\mathcal{B}}) =
(\Phi_0|_{\partial\mathcal{B}},0)$, the $s$-th order corner
condition becomes an equation of the form
\begin{equation}\label{meven}
M_0^{AB}{}_{ij}\;n_A n_B\;
\partial_n^{s} \dot{\Phi}^j(0)|_{\partial\mathcal{B}} \,\hat{=}\,
\mathrm{lower}\,\mathrm{order}
\end{equation}
for odd $s$ and
\begin{equation}\label{modd}
M_0^{AB}{}_{ij}\;n_A n_B\;
\partial_n^{s+1} \Phi^j(0)|_{\partial \mathcal{B}} \,\hat{=}\,
\mathrm{lower}\,\mathrm{order}
\end{equation}
for even $s$, where "$\hat{=}\,$lower order" means expressions which depend on lower-order,
even-numbered normal derivatives of $(\Phi(0)|_{\partial\mathcal{B}},
\dot{\Phi}(0)|_{\partial\mathcal{B}})$, modulo terms which depend on normal derivatives of the same
order, but which are zero when\\
$(\Phi(0)|_{\partial\mathcal{B}},\dot{\Phi}(0)|_{\partial\mathcal{B}})=
(\Phi_0|_{\partial\mathcal{B}},0)$. The equations
(\ref{meven},\ref{modd}) are identically satisfied if
$(\Phi(0)|_{\partial\mathcal{B}},
\dot{\Phi}(0)|_{\partial\mathcal{B}})$ and their normal
derivatives appearing on both sides are replaced by those of
$(\Phi_0|_{\partial\mathcal{B}},0)$. Using the fact that
$M_0^{AB}{}_{ij}\;n_A n_B$ depends only on first derivatives of
$\Phi$, we are now able to recursively solve the corner
conditions, with $\Phi(0)|_{\partial\mathcal{B}}$ and
$\dot{\Phi}(0)|_{\partial\mathcal{B}}$ and their even normal
derivatives given arbitrarily, provided that
$(\Phi(0)|_{\partial\mathcal{B}},
\dot{\Phi}(0)|_{\partial\mathcal{B}})$ is close to
$(\Phi_0|_{\partial\mathcal{B}}, 0)$. We have to - and by the
above can - choose a large class of initial data
$(\Phi(0),\dot{\Phi}(0))$ close to $(\Phi_0,0)$ so that the corner
conditions are fulfilled for arbitrary order and the negativity of
$M^{tt}{}_{ij}$ and the coerciveness of $M^{AB}{}_{ij}$ are
satisfied.

\end{document}